\documentstyle[preprint,floats,prd,aps]{revtex}

\def\PRref#1&#2&#3(#4){\unskip\ #1~\bf #2\rm, #3 (#4)}

\def\etal{{\it et al.}}

\begin{document}
\preprint{\vbox{\hbox{UH 511-814-95 \hfill}
                \hbox{BELLE-46 \hfill}
                \hbox{January 1, 1995 \hfill}}}

\title{A Comment on
the Experimental Determination of $|V_{ts}/V_{td}|^2$.}
\author{ T.E. Browder and S. Pakvasa}
\address{University of Hawaii at Manoa, Honolulu, Hawaii 96822}
\maketitle
\begin{abstract}
We propose a
method to extract the ratio $|V_{ts}/V_{td}|^2$
from a measurement of $\Delta \Gamma/\Gamma$ for the $B_s$ meson.
This method is experimentally
more sensitive than the conventional method for
large values of $|V_{ts}|$ but depends on the
accuracy of parton level calculations.
\end{abstract}
\bigskip

The measurement of the mixing parameter $x_s={{\Delta m}/{\Gamma}}$ for
the $B_s$ meson is one of the goals of high energy collider experiments
and experiments planned for the facilities of the future\cite{BELLE},
\cite{BABAR}.
A measurement of $x_s$ combined with a determination of $x_d$
the corresponding quantity for the $B_d$ meson allows the determination
of the ratio of the KM matrix elements ${|V_{td}|^2}/{|V_{ts}|^2}$
from the ratio $${{x_s}\over {x_d}} =
{{(m_{B_s} \eta_{QCD}^{B_s} B f_{B_s}^2)}\over
{(m_{B_d} \eta_{QCD}^{B_d} B f_{B_d}^2)}} |{{V_{ts}^2}\over {V_{td}^2}}|$$
The factor which multiplies the ratio of KM matrix elements
is estimated to be unity up to $SU(3)$ breaking effects and has
been estimated to be of order $(1.3)$\cite{Forty}.
Since time integrated measurements
of $B_s$ mixing are insensitive to $x_s$ when mixing is maximal,
 one must make time dependent measurements in order
to extract this parameter.
A severe experimental difficulty
is the rapid oscillation rate of the $B_s$ meson, as recent experimental
limits indicate that $x_s > 8.5$\cite{Forty} and theoretical fits to
the Standard Model parameters suggest that $x_s$ lies in the
range $10-40$.

It should be noted that there is another parameter of the $B_s$ meson
which can also be measured, this is $\Delta \Gamma/\Gamma$, the difference
between the widths of the two $B_s$ eigenstates.
For $V_{ts}\sim 0.043$
this could lead to a value of $\Delta \Gamma/\Gamma$ of order
$10-20\%$ which is measurable at high energy experiments or
asymmetric B factories.
In parton calculations\cite{Hagelin}
$$ \Delta\Gamma = {{-G_F^2 f_B^2 m_B m_b^2 \lambda_t^2} \over {4 \pi}}
{}~[1+ {4\over 3}{{\lambda_c}\over{\lambda_t}} {{m_c^2} \over {m_b^2}} +
O(m_c^4/m_b^4)]$$
Comparing to the dispersive term, this gives
$${{\Delta \Gamma_{B_s}} \over {\Delta m_{B_s}}}\approx {-3\over 2} \pi~
{{m_b^2}\over{m_t^2}}\times
{ {\eta_{QCD}^{\Delta \Gamma(B_s)})}\over
{\eta_{QCD}^{\Delta M(B_s)} }} $$
where $m_b$, $m_t$ are the masses of the b and t quark
respectively and terms of
order $m_c^2/m_b^2,~m_b^2/m_t^2$
are neglected\cite{Hagelin}.
The last factor in the above
expression, the ratio of QCD
corrections for $\Delta \Gamma$ and
$\Delta M$, is expected to be of order unity.
Thus the ratio ${{x_s}/ {x_d}}$ is given by,
$${{\Delta \Gamma_{B_s}} \over{\Delta m_{B_d}}} = {-3\over 2} \pi
  ~{{m_b^2}\over{m_t^2}}~
{{(m_{B_s} \eta_{QCD}^{\Delta \Gamma(B_s)} ~B f_{B_s}^2)}\over
{(m_{B_d} \eta_{QCD}^{\Delta M(B_d)}
{}~B f_{B_d}^2)}} {{|V_{ts}|^2}\over {|V_{td}|^2}}$$
All of the above factors in $\Delta\Gamma$
have a common mass dependence of $m_b^2$
in the leading term. We have also assumed that the
lifetimes of the $B_d$ and $B_s$ mesons are equal, although the
formula can be easily modified to take into account any measured
lifetime difference.
The above expression assumes unitarity since the
leading term
which enters in $\Delta\Gamma$ is $$\lambda_u^2 + \lambda_c^2 +
2 \lambda_u \lambda_c $$
which is expressed as
$$(\lambda_c + \lambda_u)^2 = (\lambda_t)^2 $$ via unitarity of the KM
matrix.
In fact, all determinations of $V_{ti}$ which depend on virtual t quarks
necessarily rely on the assumed unitarity of the $3\times 3$ KM matrix.
The only way to obtain values of $V_{ti}$ free from this assumption
is through direct on-shell measurements of t decays.

Several authors have pointed out that the quantity
$\Delta\Gamma (B_s)$ may be large since
there are intermediate final states such as $\bar{B}_s\to D_s^{(*)+}
\bar{D_s^{(*)-}}$
accesible to both $B_s$ and $\bar{B}_s$
which have appreciable branching states\cite{Aleksan}.
The calculation of Aleksan, Le Yaouanc,
Oliver, Pene, and Raynal\cite{Aleksan} shows that the
parton model estimate and the calculation using exclusive final states
agree to within an accuracy of 30\%.
Given the large experimental uncertainties
already present in the determination of the ratio $|V_{ts}/V_{td}|^2$,
this is not yet a serious limitation.
In the future, it will also be
possible to verify that the assumptions
in the parton model estimate and in the calculation of Aleksan\etal
are appropriate by
detailed measurements of $B_s$ decay rates.
For example, one may check that the W-exchange is small by searching
for $\bar{B}_s \to D^{(*)+} \bar{D}^{(*)-}$ and $\bar{B}_s \to D^+ \pi^-$
\cite{conjugate}.
Verification of the
expected pattern of branching ratios predicted in ref.~\cite{Aleksan}
should also be possible.
The required measurements would be possible at a lengthy dedicated
run on the $\Upsilon(5S)$ resonance at a B Factory.

In order to measure ${{\Delta \Gamma}/{\Gamma}}$, one must
determine the lifetimes
of two samples of events\cite{DassSarma}. One possiblity is
to use the large samples of $\bar{B}_s\to \psi\phi$ events
and $\bar{B}_s\to D_s^{(*)} ~l~\nu$ events.
The first sample may be dominated by
events in a single CP eigenstate as is the case for $B_{d,u}\to \psi K^*$.
This can be verified experimentally by measuring the polarization in
this decay. The latter sample of semileptonic decays will be an incoherent
mixture of both CP eigenstates. The measured lifetime difference
will be $\Delta \Gamma/\Gamma^2$, which can then be used to constrain
$|V_{ts}|^2/|V_{td}|^2$.

The sensitivity of the two methods can be roughly compared as follows.
The ALEPH lower limit on $x_s$ (8.5) corresponds to the lower limit
 $\Delta\Gamma/\Gamma$ $>0.033$~(3.3\%). A measurement of
a $7\%$ lifetime difference corresponds to a central value of
$x_s= 15$ for a time dependent oscillation study.
For large values of $V_{ts}$, the method using $\Delta \Gamma$
eventually becomes more sensitive.
Good control of systematic effects from the
boost correction in $\bar{B}_s\to D_s^+ \ell^- \nu$ and the lifetime of
the background sample are required. In published lifetime analyses,
these effects have been estimated at the 3-7\% level\cite{Forty}.
However, systematic uncertainties in high statistics measurements
of charm meson lifetimes have been
sucessfully reduced to the 1\% level\cite{Forty}. Similar improvements
with higher statistics can be anticipated for B hadron lifetime
measurements.

The disadvantage of this method of constraining the KM matrix (using
$|\Delta\Gamma_{B_s}|$ and $|\Delta m_{B_d}|$) is the greater
reliance on parton calculations. In addition to the methods described
here which involve $B_s-\bar{B_s}$ mixing, constraints on $|V_{ts}|^2/
|V_{td}|^2$ can be obtained by comparing
the branching fractions of  $B\to K^{*} \gamma$ and
$B\to \rho\gamma$ decays.
As the precision of these measurements of rare decays
improves, a careful treatment of possible long distance contributions
and $SU(3)$ breaking effects will be required
\cite{Golowich},\cite{Cheng},\cite{atwood}.

Finally, it is expected that the contribution of $\Delta \Gamma$ to
$B_d-\bar{B_d}$ mixing is small since there are no final states
that are accessible to both $B_d$ and $\bar{B_d}$ mesons that
also have appreciable
branching ratios.
This plausible assumption should eventually
be experimentally
verified by comparing lifetime measurements of $B_d$ in CP eigenstates
such as $\bar{B}_d\to \psi K^0$
and $\bar{B}_d\to D^{*+} \ell^- \nu$. The present
experimental precision allows for a substantial $\Delta\Gamma$
 contribution to the observed time integrated $B_d$ mixing rate.


\begin{references}
%
\bibitem{BELLE}
{K.\ Abe et al., {\em Physics and Detector of Asymmetric B Factory
at KEK}, KEK Report 90-23 (1991). Also see {\it Letter of Intent
for A Study of CP Violation in B Meson Decays}, KEK Report 94-2, April 1994.}
\bibitem{BABAR}
{Letter of intent for the Study of CP Violation and Heavy Flavor Physics
at PEPII, The BABAR collaboration, SLAC report SLAC-443, 1994.}
\bibitem{Forty}{R. Forty, CERN preprint CERN-PPE/94-154, to appear
in the Proceedings of the ICHEP94 Conference, Glasgow, Scotland.}
\bibitem{Hagelin}{J. Hagelin, Nucl Phys. B 193 (1981), 123.
See also A.J. Buras, W. Stominiski,
and H. Steger, Nucl Phys. B 245 (1984), 369;
M.B. Voloshin, N.G. Uraltsev, V.A. Khoze, and
M.A. Shifman, Sov. J. Nucl. Phys. {\bf 46}, 112, 1987.}
\bibitem{Aleksan}{R. Aleksan, A. Le Yaouanc, L. Oliver, O. Pene, and
J.C. Raynal, Phys. Lett. B 316, 567 (1993).}
\bibitem{conjugate}{Unless otherwise noted, the charge conjugate
reaction is implied.}
\bibitem{DassSarma}
{Dass and Sarma have suggested another method of determining $\Delta \Gamma$
from measurements of time dependent asymmetries of
lepton tagged neutral B meson decays
to CP eigenstates\cite{Sarma}.
They show that the asymmetry is given by $\pm \rm{tanh}(\Delta \Gamma \tau)$
where the $\pm$ sign refers to the CP eigenvalue.
This will allow the resolution of the
four fold ambiguity and determine which is more massive of the short or long
lived eigenstates. This may be feasible at B factories.}
\bibitem{Sarma}{G.V. Dass and K.V.L. Sarma, Int. Journal of Modern Phys.,
24, 6081 (1994).}
\bibitem{Golowich}{E. Golowich and S. Pakvasa, UH-511-800-94, UMHEP-411,
to appear in Physical Review D.}
\bibitem{Cheng}{H.Y. Cheng, Academica Sinica preprint IP-ASTP-23-94.}
\bibitem{atwood}{D. Atwood, B. Blok, and A. Soni,
SLAC-PUB-6635, BNL-60709, TECHNION-PH-94-11.}
\end{references}
\end{document}